\documentclass[11pt]{article}
\usepackage{geometry}                % See geometry.pdf to learn the layout options. There are lots.
\usepackage{graphicx}
\usepackage{amssymb}
\usepackage{epstopdf}
\usepackage{graphicx}

\usepackage{float}

\usepackage{color}
\usepackage{times}

\usepackage[normalem]{ulem}

\usepackage[textsize=footnotesize]{todonotes}

\def \ii {\mbox{i}}

\newcommand{\be}{\begin{equation}}
\newcommand{\ee}{\end{equation}}

\newcommand{\bea}{\begin{eqnarray}}
\newcommand{\eea}{\end{eqnarray}}

\usepackage{bm}

\title{Explosive Synchronization is Discontinuous}
\author{Vladimir Vlasov$^1$,  Yong Zou$^2$ , and  Tiago Pereira$^3$ \\ Institute of Physics, Potsdam University, Germany \\  Department of Physics, East China Normal University, Shanghai,
200062, China \\ 
Department of Mathematics, Imperial College London, London SW72AZ, UK}
	
\date{\today}

\begin{document}
\maketitle
%\section{}
%\subsection{}

\begin{abstract}
Spontaneous explosive is an abrupt transition to collective behavior taking place in heterogeneous networks when the frequencies of the nodes are positively correlated to the node degree. This explosive transition was conjectured to be discontinuous. Indeed,  numerical investigations reveal a hysteresis behavior associated with the transition. Here, we analyze explosive synchronization in star graphs. We show that in the thermodynamic limit the transition to (and out) collective behavior is indeed discontinuous. The discontinuous nature of the transition is related to the nonlinear behavior of the order parameter, which in the thermodynamic limit exhibits multiple fixed points. Moreover, we unravel the hysteresis behavior in terms of the graph parameters. Our numerical results show that finite size graphs are well described by our predictions.
\end{abstract}

%\pacs{05.45.Xt, 89.75.Hc, 05.45.Ac}

\section{Introduction}

Many networks in nature exhibit a heterogeneous behavior in the node's degrees, with some nodes being massively connected whereas the remaining nodes are sparsely connected \cite{Barrat}. These heterogeneous networks exhibit intricate collective properties \cite{Takashi,Lee,HubSync}. Of particular interest is a situation where on top of each node we have an oscillator whose (isolated) frequency positively correlates with the node's degree.  In such a setting, as one increases the interaction strength among nodes an abrupt transition towards collective behavior is observed \cite{JesusPRL2011}. This abrupt transition to collective behavior,  called explosive synchronization,  appears not to be model specific and has been observed  in electronic circuits \cite{LeyvaPRL2012}, time delayed systems \cite{PeronPRE2012}, and a second order Kuramoto model~\cite{Ji_PRL2013}.

Two distinct features of explosive synchronization have attracted a great deal of attention. First, its abrupt character leading to an apparently discontinuous transition \cite{JesusPRL2011,LeyvaPRL2012,PeronPRE2012,Ji_PRL2013,Skardal_EPL2013,Li_PRE2013,Wuye_EPL2013,Leyva_SciRep2013,Coutinho_PRE2013,Peron_PhysRevE2012}. Second, the hysteresis behavior associated with the two distinct transitions: forward and backward transitions. The forward transition is characterized by  a fast transition from an incoherent state to a coherent one as the interaction strength increases. Whereas in the backward transition one starts at a stable coherent state and as the coupling strength decreases a sudden transition to an incoherent state takes place. The critical coupling for the forward transition does not coincident with that of the backward transition. Moreover, there are evidence that such a hysteretic loop is determined by the networks structural parameters \cite{Yong_PRL2013}. 

To uncover the nature of the transition, one needs to understand the global behavior of the order parameter. However, this requires one to globally solve the equations of motion, which is impractical for a general network. Recently, using star graphs as motifs and a series of approximations it was possible to study the transition and thereby to reinforce the abrupt character of the transition \cite{Coutinho_PRE2013,Peron_PhysRevE2012,Yong_PRL2013,Connectivity}. However,  the dynamical behavior of the order parameters and the precise nature of the transition remain undisclosed.

We investigate explosive transition of Kuramoto oscillators in a star graph, which consists of a central hub connected to $N$ nodes. Here, we show that in the thermodynamics limit, when the system size goes to infinity, the model is fully solvable. We obtain two fundamental results:
\begin{enumerate}
\item Transitions associated with explosive synchronization are discontinuous.
\item There is a hysteresis loop associated with explosive synchronization. 
\end{enumerate}
We obtain these results by employing the Wattabe-Strogatz (WS) approach \cite{Watanabe-Strogatz-93,Watanabe-Strogatz-94,Pikovsky-Rosenblum-11}, which yields an exact nonlinear equation for the order parameter. The stability analysis reveals the existence of parameter regions where the order parameter exhibits multi-stability. These coexisting fixed points lead to the hysteresis behavior and to both forward and backward discontinuous transition. The main mechanism generating the forward discontinuity is the following
\begin{center}
\begin{minipage}[c]{0.9\textwidth}
The asynchronous dynamics corresponds to a fixed point of the order parameter. After reaching a critical coupling value the fixed point disappears and the only remaining fixed point  corresponds to the synchronized state. 
\end{minipage}
\end{center}
We also check our predictions against finite size effects. Our numerics reveals that the finite size fluctuations are negligible.  Since, star graphs are fundamental building blocks for intricate structures such as scale-free networks, our findings can be used as paradigms to unravel the transition.

\section{The Model}

Let us consider the star network of $N$ leave nodes and a central hub described by 
\begin{eqnarray}\label{mod1}
\dot{\varphi_k}&=&\omega+\lambda \sin ( \psi - \varphi_k) \,\,\,\,\,\,\,\,\, \mbox{ for  } k=1,2,\cdots,N   \\
\dot{\psi}&=& \beta \omega+\beta  \frac{\lambda}{N} \sum_{k=1}^N \sin (\varphi_k - \psi) 
\end{eqnarray}
where we have normalized the frequency of the leaves to unit and $\beta >1$ corresponding to the positive correlation between  the degree and frequency. Here, we introduce a slight modification to Refs. \cite{JesusPRL2011,Yong_PRL2013}, which concerns the  rescaled coupling strength for the hubs. The reason is that we will consider the thermodynamic limit $N\rightarrow \infty$, so the normalization is necessary  to make sense of the limit. In practice, this normalization is immaterial as one only considers finite size effects.

\section{Fully Solvable Limit: Nonlinear evolution law for the order parameter}

Our goal is to employ the WS change of coordinates and solve the model fully. The model is fully solvable because all nodes have the same frequency (isolated).  To perform the analysis, we first introduce the phase difference
\begin{equation}	\label{var.1}
\theta_k=\varphi_k-\psi.
\end{equation}

Next, we rewrite the equations for $\theta_k$ in the setting the of WS approach
\begin{eqnarray} \label{id.1}
\dot{\theta_k}&=&-(\beta-1)\omega- \beta \lambda {\rm Im}(G(t))+\lambda{\rm Im}(e^{-\ii \theta_k}),\\
G(t)&=&\frac{1}{N}\sum_{j=1}^N e^{\ii\theta_j},
\end{eqnarray}
and the equation for $\psi$ remains unchanged. Now we applied the WS approach for Eq.~(\ref{id.1}) ~\cite{Watanabe-Strogatz-93,Watanabe-Strogatz-94}.
We make use of the WS variable transform in the form presented in~\cite{Pikovsky-Rosenblum-11}, that is, 
\begin{equation}\label{tr.WS.1}
e^{\ii\varphi_k}=\frac{z+e^{\ii(\xi_k+\alpha)}}{1+z^*e^{\ii(\xi_k+\alpha)}},
\end{equation}
where $z=z(t)$, $\alpha=\alpha(t)$ are global variables and $\xi_k$ are constants that depend on the initial conditions of the system~(\ref{id.1}) (for more details see~\cite{Pikovsky-Rosenblum-11}). Next, we need to express $G(t)$ in terms of new variables. In the general case, the expression for $G(t)$ is rather complex, but in the thermodynamic limit $N\to\infty$ and uniform distribution of constants $\xi_k$
$$
\sigma(\xi)=(2\pi)^{-1},
$$
where $\sigma$ denotes the density. In this variables, we obtain
$$
G(t)= z,
$$ 
where $z$ is a complex number and $|z|$ is the order parameter. The WS approach yields a closed system of equations for the global macroscopic variables 
\begin{eqnarray}\label{id.WS.1}
\displaystyle \dot{z}&=& \ii[-(\beta-1)\omega-\beta \lambda{\rm Im}(z)]z+\lambda \frac{1-z^2}{2}, \\
\dot{\alpha}&=&-(\beta-1)\omega-\beta \lambda{\rm Im}(z) +\lambda{\rm Im}(z^*).
\end{eqnarray}
Since $\alpha$ does not enter in the equation for $z$, so we can consider the equation for $z$ separately.
	
\subsection{Fixed Points for the Order parameter}	
	
\begin{figure}[h]
\begin{center}
\includegraphics[width=0.55\columnwidth]{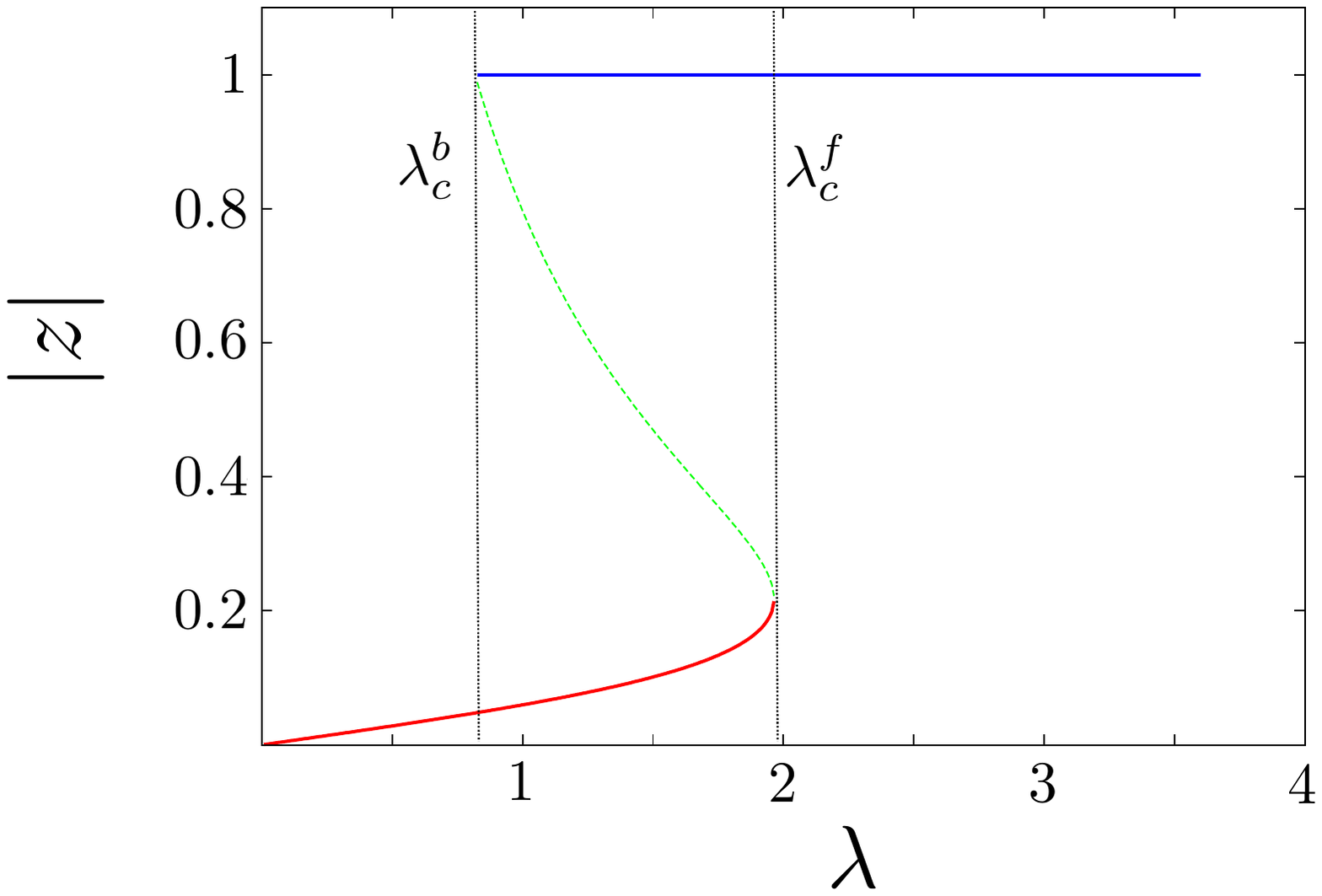}
\caption{Dependence of the order parameter $|z|$ on the coupling $\lambda$. We show the critical forward and backward couplings for $\omega=1$, $\beta=10$. The red line shows the fixed point ${z_a}_{2}$ corresponding to the incoherent state. The green dashed line shows the fixed point with ${z_a}_{1}$ corresponding to the separatrix between coherent and incoherent states. Finally, the blue line shows the fixed points ${z_s}_{1,2}$ with $|z_s|=1$, which correspond to the coherent state. }
\label{fig.id.2.fp}
\end{center}
\end{figure}

The fixed points of the first equation in~(\ref{id.WS.1}) are
\begin{eqnarray}\label{id.2.fp.1}
\displaystyle {z_s}_{1,2}=\exp \left\{ \ii \arcsin\left(-\frac{(\beta-1)\omega}{(\beta+1)\lambda}\right) \right\}, \ \ \ &\mbox{if}\ \lambda> \lambda_c^f,\\
~ \nonumber \\
\displaystyle {z_a}_{1,2}=-\ii\,\frac{(\beta-1)\omega\pm\sqrt{(\beta-1)^2\omega^2-(2\beta+1)\lambda^2}}{(2\beta+1)\lambda}, \ \ \ &\mbox{and} \\
~ \nonumber \\
\displaystyle {z_s}_{1,2}=\exp \left\{\ii \arcsin\left(-\frac{(\beta-1)\omega}{(\beta+1)\lambda}\right) \right\}, \ \ \ &\mbox{if}\ \lambda_c^f \geq\lambda \geq \lambda_c^b, \\
~ \nonumber \\
\displaystyle {z_a}_{2}=-\ii\,\frac{(\beta-1)\omega-\sqrt{(\beta-1)^2\omega^2-(2\beta+1)\lambda^2}}{(2\beta+1)\lambda}, \ \ \ &\mbox{if}\ \lambda<\lambda_c^b.
\end{eqnarray}
where $\lambda_c^f$ and $\lambda_c^b$ corresponds to the forward and backward critical couplings
$$
\lambda_c^f  = \frac{\beta-1}{\sqrt{2\beta+1}}\omega \,\,\,\, \mbox{  and    } \,\,\,\, \lambda_c^b = \frac{\beta-1}{\beta+1}\omega \,\,\,\,
$$
In order to study the stability of the fixed points~(\ref{id.2.fp.1}) we linearize the system around the corresponding fixed point. First, for two asynchronous fixed points ${z_a}_{1,2}$ we obtain the following linearized system:
\begin{eqnarray} \label{id.2.stab.asyn}
\displaystyle \dot{a}&=&\mp\left((\beta+1)\sqrt{\lambda_c^b-\frac{2\beta+1}{(\beta+1)^2}\lambda^2\,}\,\right)b, \nonumber \\
&~& \nonumber\\
\displaystyle \dot{b}&=&-\left(\frac{\beta(\beta+1)}{2\beta+1}\lambda_c^b \mp\frac{(\beta+1)^2}{2\beta+1}\sqrt{\lambda_c^b-\frac{2\beta+1}{(\beta+1)^2}\lambda^2\,}\,\right)a, \nonumber
\end{eqnarray}
where $a={\rm Re}(z)$ and $b={\rm Im}(z)-{\rm Im}({z_a}_{1,2})$ respectively. From (\ref{id.2.stab.asyn}) it follows that the fixed point with ``+" ${z_a}_{1}$ is a saddle for $\lambda>\lambda_c^b$ and the fixed point with ``-" ${z_a}_{2}$ is neutrally stable center (for more detail see~\cite{Vlasov-Pikovsky-Macau-unpublished}). Second, for two synchronous fixed points ${z_s}_{1}$ (with positive real value) and ${z_s}_{2}$ (with negative real value):
\begin{eqnarray} \label{id.2.stab.syn}
\displaystyle \dot{a}&=&\mp\left(\sqrt{\lambda^2-(\lambda_c^b)^2\,}\,\right)a- \lambda_c^b b, \nonumber \\
&~& \nonumber\\
\displaystyle \dot{b}&=&\mp\left((\beta+1)\sqrt{\lambda^2-(\lambda_c^b)^2\,}\,\right)b, \nonumber
\end{eqnarray}
where $a={\rm Re}(z)-{\rm Re}({z_s}_{1,2})$ and $b={\rm Im}(z)-{\rm Im}({z_s}_{1,2})$ respectively. From (\ref{id.2.stab.syn}) it follows that for $\lambda>\lambda_b^c$ the fixed point with positive real value ${z_s}_{1}$ is sink (stable) node and the fixed point with negative real value ${z_s}_{2}$ is source (unstable) node. The fixed points~(\ref{id.2.fp.1}) are shown on (Fig.~\ref{fig.id.2.fp}) for $\omega=1$, $\beta=10$.
We recap the  stability of the fixed points and physical meaning for our problem in Table \ref{tab1}. 
\begin{table}[ht]
\caption{Fixed points of the order parameters with their stability and meaning.}
\centering
\begin{tabular}{c c c c} % Column formatting, @{} suppresses leading/trailing space
\hline\hline
Fixed Point & Stability & Existence Region & Physical Meaning \\
\hline
\hline
${z_s}_{1}$ & sink  &  $\lambda > \lambda_c^b$ & Coherent State \\
${z_s}_{2}$ & source  & $\lambda > \lambda_c^b$ & none \\ 
${z_a}_{1}$ & saddle & $\lambda_c^b > \lambda > \lambda_c^f$ & separatrix \\
${z_a}_{2}$ & center & $\lambda > \lambda_c^f$ &  Incoherent State\\
\hline
\end{tabular}
\label{tab1}
\end{table}

\section{Numerics}

Although the non-linear equation (\ref{id.WS.1}) for the order parameter is obtained in the thermodynamical limit, our numerical experiments reveal that finite size effects are negligible. Our predictions are in excellent agreement with the numerical experiments. We discuss this situation in some detail.

\begin{figure}[H]
\begin{center}
\includegraphics[width=1\columnwidth]{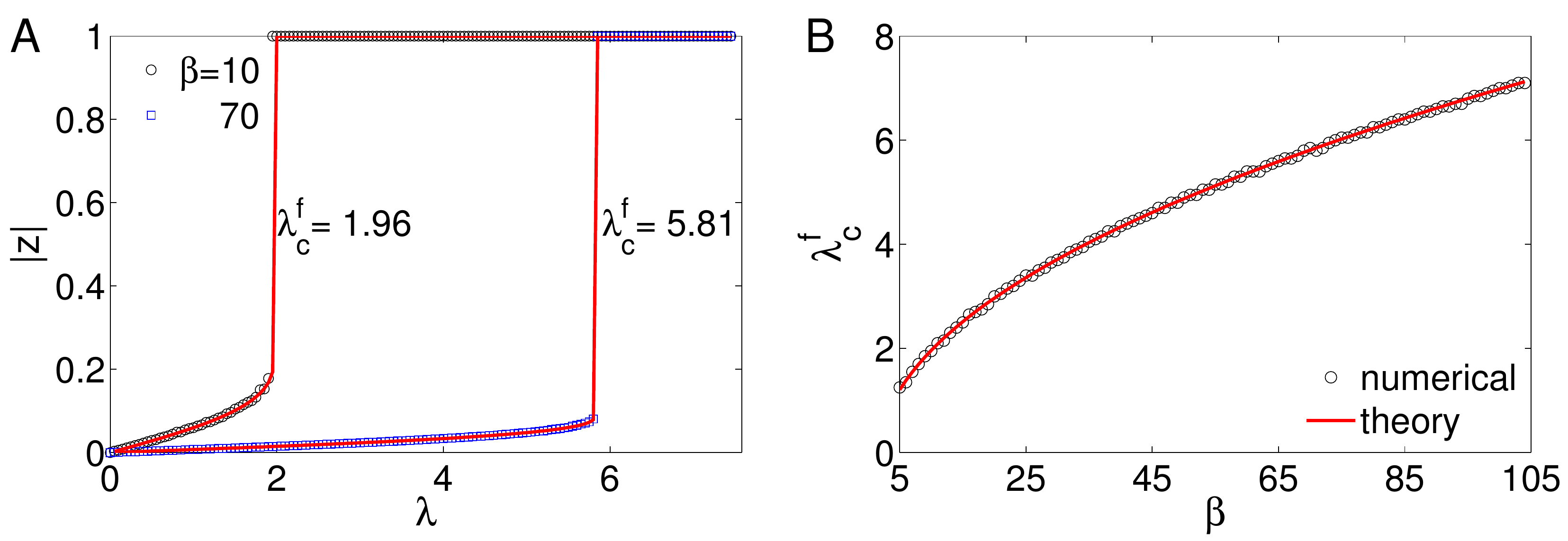}
\caption{Numerical simulations of the forward transition for $N=10000$. (A) Order parameter $|z|$ versus the
coupling strength $\lambda$ obtained numerically  for $\beta = 10$ and  $\beta = 70$. The full lines are the
thermodynamical predictions ${z_a}_2$. (B) The critical forward
coupling $\lambda_c^f$ versus the $\beta$. The numerical values
are shown as open circles, while the (red) thick line is the
theoretical prediction.}
\label{numerics}
\end{center}
\end{figure}

We have numerically simulated the model (\ref{mod1}) using a fourth order Runge-Kutta integrator. We have focused on the forward transition. The experiments are performed as follows: For a fixed value of $N$ and $\beta$, we start the simulations with $\lambda=0$ and uniformly distributed initial conditions, then we integrate the model up to a time $T$. Then we increase the value of the coupling $\lambda$ by a small amount and use the final step of the previous simulation as initial conditions. This procedure is known as following the attractor. Then for each step of this procedure we compute the order parameter $|z|$ as described in (\ref{id.1}). Since we are computing the order parameter corresponding to the incoherent state, we check our numerics against the thermodynamic limit described by ${z_a}_{2}.$ Our numerics shows that the agreement is excellent, see Fig. \ref{numerics}.
%{\color{red} Should we add a small note on uniform IC for each coupling? This is
%a special case for this model since $|z|_{a_2}$ is neutrally stable. }

\section{Conclusions}

We have analyzed explosive synchronization for star graphs. We demonstrated that in the limit of infinitely many leave nodes the order parameter is governed by a nonlinear equation, which we were fully able to analyze.  We showed the transitions towards and out of coherence are discontinuous. Moreover, we revealed the mechanism of discontinuity, namely, bifurcations of the order parameter create and annihilate fixed points forcing the loss of incoherence states (likewise the incoherent state).

In our model the parameter $\beta$ controls the positive correlation between structure and dynamics. Our analysis shows that any positive correlation $\beta > 1$ leads to discontinuous transitions. Although, our macroscopic equations for the order parameters are particular to the star graphs, these graphs can be used as motifs to more intricate structures such as scale-free networks, see the discussion in Ref. \cite{Yong_PRL2013}. Therefore, our results also shed light into the transitions in complex networks where the star motif is dominant.

{\bf Acknowledgments}: This work was partially supported by DFG/FAPESP grant IRTG 1740/TRP 2011/50151-0. YZ was supported by NNSFC (Grant Nos. 11305062, 11135001, 81471651) (YZ) and TP by the Marie Curie IIF Fellowship (Project 303180).

%----------------------------------------------------------------------------

\end{document}